\documentclass[apjl]{emulateapj}
\usepackage{color,lineno,url}

\shorttitle{First {\it NuSTAR} solar AR observations.}

\shortauthors{Hannah et al.}

\begin{document}

\title{The first X-ray imaging spectroscopy of \\quiescent solar active regions with {\it NuSTAR}}
\author{Iain G. Hannah\altaffilmark{1}, Brian W. Grefenstette\altaffilmark{2}, David M. Smith\altaffilmark{3}, Lindsay Glesener\altaffilmark{4,5}, S{\"a}m Krucker\altaffilmark{5,6}, Hugh S. Hudson\altaffilmark{1,5}, Kristin K. Madsen\altaffilmark{2}, Andrew Marsh\altaffilmark{3}, Stephen M. White\altaffilmark{7}, Amir Caspi\altaffilmark{8}, Albert Y. Shih\altaffilmark{9},
Fiona A. Harrison\altaffilmark{2}, Daniel Stern\altaffilmark{10}, 
Steven E. Boggs\altaffilmark{5}, Finn E. Christensen\altaffilmark{11}, William W. Craig\altaffilmark{5,12}, Charles J. Hailey\altaffilmark{13}, William W. Zhang\altaffilmark{14}}

\altaffiltext{1}{SUPA School of
Physics \& Astronomy, University of Glasgow, Glasgow G12 8QQ,
UK}\email{iain.hannah@glasgow.ac.uk}
\altaffiltext{2}{Cahill Center for Astrophysics, 1216 E. California Blvd, California Institute of Technology, Pasadena, CA 91125, USA}
\altaffiltext{3}{Santa Cruz Institute of Particle Physics and Department of Physics, University of California, Santa Cruz, CA 95064, USA}
\altaffiltext{4}{School of Physics \& Astronomy, University of Minnesota - Twin Cities, Minneapolis, MN, 55455, USA}
\altaffiltext{5}{Space Sciences Laboratory University of California, Berkeley, CA 94720, USA}
\altaffiltext{6}{Institute of 4D Technologies, School of Engineering, University of Applied Sciences and Arts Northwestern Switzerland, 5210 Windisch, Switzerland}
\altaffiltext{7}{Air Force Research Laboratory, Space Vehicles Directorate, 3550 Aberdeen Ave SE, Kirtland AFB, NM 87117, USA}
\altaffiltext{8}{Southwest Research Institute, Boulder, CO 80302, USA}
\altaffiltext{9}{Solar Physics Laboratory, NASA Goddard Space Flight Center, Greenbelt, MD 20771, USA}
\altaffiltext{10}{Jet Propulsion Laboratory, California Institute of Technology, Pasadena, CA 91109, USA}
\altaffiltext{11}{DTU Space, National Space Institute, Technical University of Denmark, Elektrovej 327, DK-2800 Lyngby, Denmark}
\altaffiltext{12}{Lawrence Livermore National Laboratory, Livermore, CA 94550, USA}
\altaffiltext{13}{Columbia Astrophysics Laboratory, Columbia University, New York, NY 10027, USA}
\altaffiltext{14}{Astrophysics Science Division, NASA Goddard Space Flight Center, Greenbelt, MD 20771, USA}

\begin{abstract}
We present the first observations of quiescent active regions (ARs) using {\it NuSTAR}, a focusing hard X-ray telescope capable of studying faint solar emission from high temperature and non-thermal sources. We analyze the first directly imaged and spectrally resolved X-rays above 2~keV from non-flaring ARs, observed near the west limb on 2014 November 1. The {\it NuSTAR} X-ray images match bright features seen in extreme ultraviolet and soft X-rays. The {\it NuSTAR} imaging spectroscopy is consistent with isothermal emission of temperatures $3.1-4.4$~MK and emission measures $1-8\times 10^{46}$~cm$^{-3}$. We do not observe emission above 5~MK but our short effective exposure times restrict the spectral dynamic range. With few counts above 6~keV, we can place constraints on the presence of an additional hotter component between 5 and 12~MK of $\sim 10^{46}$cm$^{-3}$ and $\sim 10^{43}$ cm$^{-3}$, respectively, at least an order of magnitude stricter than previous limits. With longer duration observations and a weakening solar cycle (resulting in an increased livetime), future {\it NuSTAR}  observations will have sensitivity to a wider range of temperatures as well as possible non-thermal emission.
\end{abstract}


\section{Introduction}

The detailed process by which energy is released in the Sun's atmosphere remains poorly understood. In active regions (ARs), the heating of material and acceleration of particles occur impulsively in flares, but even when there appear to be no flares, ARs still contain loops of material heated to several million Kelvin. One proposed solution is to have episodic heating, with events frequent enough to smooth out the time series \citep[e.g.][]{1975ApJ...199L..53G}. \citet{1988ApJ...330..474P} described this explanation in magnetohydrodynamic terms as the natural development of small current sheets in magnetic flux tubes that appear as coronal loops, and coined the term ``nanoflare'' to represent the idea, estimating that an individual event might contain $\sim 10^{-9}$ the energy of a major flare. This coronal energy release would be driven by the movement of the magnetic loops' footpoints. These need to be sufficiently slow (longer than the Alfv{\'e}n time for wave propagation) otherwise waves would dominate and be an alternative heating mechanism \citep[e.g.][]{2006SoPh..234...41K,2014LRSP...11....4R}.

\begin{figure*}
\centering
\includegraphics[width=0.89\columnwidth]{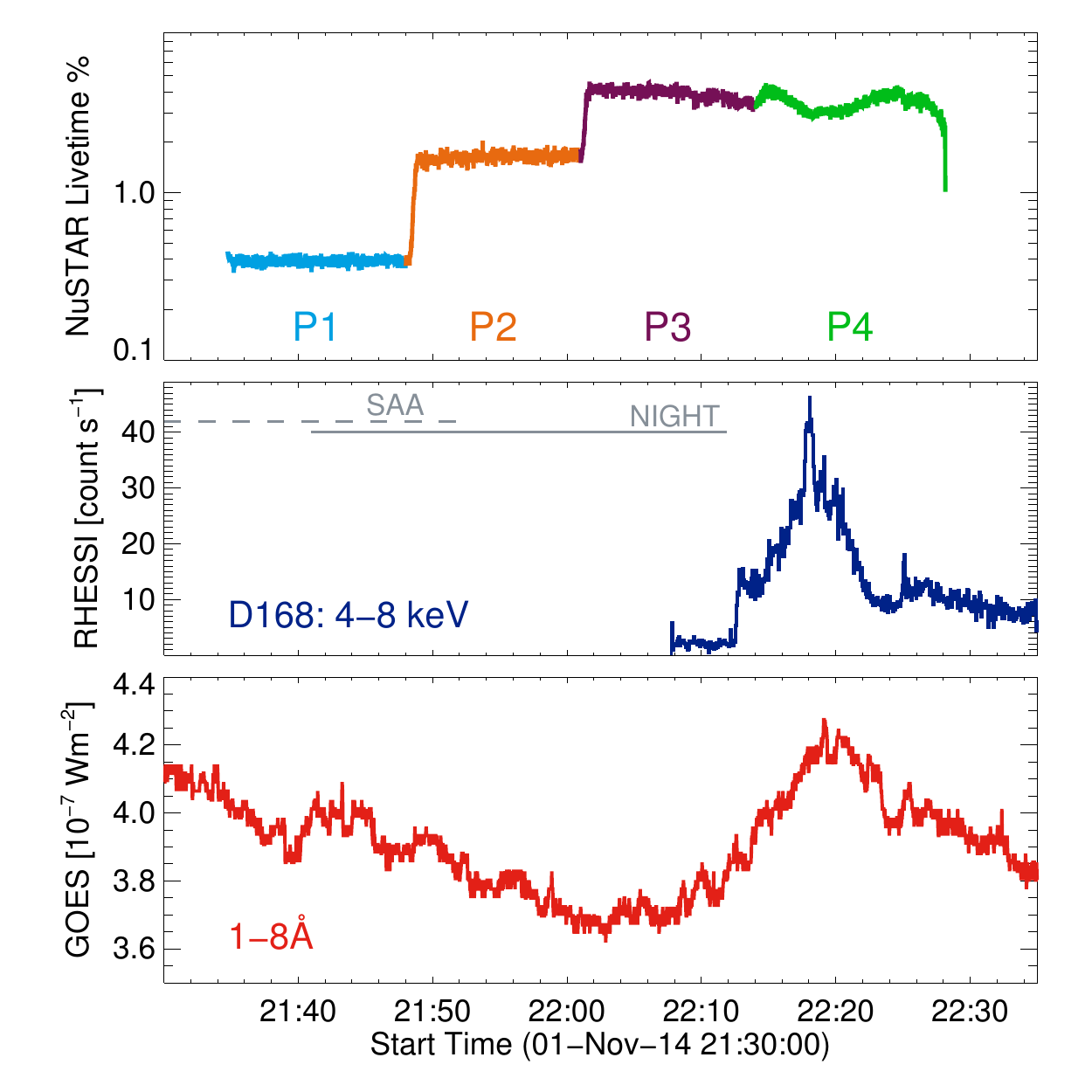}
\includegraphics[width=0.95\columnwidth]{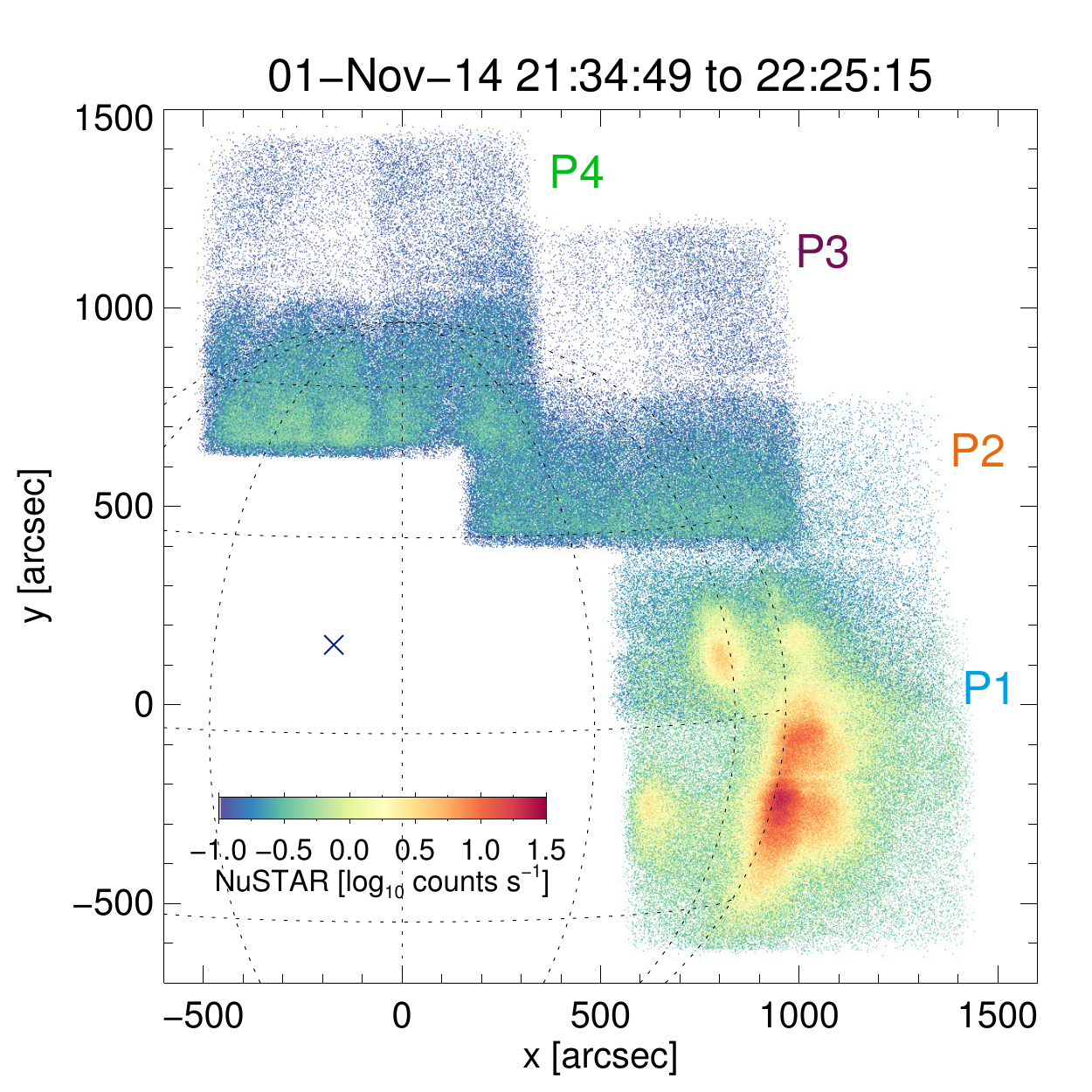}
\caption{(Left) Time profile of the {\it NuSTAR} livetime for the four limb pointing positions (top), the {\it RHESSI} 4-8~keV count rate (middle) and the {\it GOES} 1-8~\AA~flux (bottom). (Right) {\it NuSTAR} image $2-78$~keV of the four limb pointings combined together, relative to Sun center. The `x' indicates the 22:18UT microflare.}
\label{fig:lght_mos}
\end{figure*}

The nanoflare conjecture implies that the plasma temperature temporarily, and locally, will exceed the mean \citep{1990ApJ...356L..31S,1994ApJ...422..381C,2014ApJ...784...49C,2014LRSP...11....4R}. Individual nanoflares are likely to be so small that they will be difficult to detect, but their presence should be observable through the impact of an unresolved ensemble. The spectrum would contain a range of temperatures reflecting an assortment of nanoflares at different energies and stages of their energy redistribution. Extreme ultraviolet (EUV) observations with {\it SDO}/AIA and {\it Hinode}/EIS of non-flaring ARs typically show a Differential Emission Measure (DEM) peaked about 3~MK, steeply falling off to higher and lower temperatures \citep[e.g.][]{2011ApJ...734...90W,2012ApJ...759..141W,2015A&A...573A.104D}. The slope of the DEM is well studied over $1-3$~MK due to the number of EUV lines observable from quiescent regions, giving scalings of $EM\approx T^{3-5}$. This is consistent with high-frequency events, where the time between impulsive heatings is shorter than the cooling timescale \citep{2004ApJ...605..911C,2014LRSP...11....4R}. The higher temperature ($>$5~MK) slope has considerably larger uncertainties, due to weaker EUV diagnostics. It appears to be steeper, with $EM\approx T^{-(6-10)}$ \citep{2012ApJ...759..141W}. Observed lines include \ion{Fe}{18} \citep[{\it SUMER};][]{2012ApJ...754L..40T} and \ion{Fe}{19} \citep[{\it EUNIS};][]{2014ApJ...790..112B}, with peak formation temperatures of 7.1~MK and 8.9~MK respectively. {\it SDO}/AIA 94\AA~can also be used to access \ion{Fe}{18} \citep{2011ApJ...736L..16R,2011ApJ...734...90W,2012ApJ...750L..10T,2012ApJ...759..141W,2013A&A...558A..73D}.

The hottest material would produce X-ray emission but is difficult to detect given the expected weak signal and instruments designed for brighter flares. {\it Hinode}/XRT has been used with {\it Hinode}/EIS \citep{2011ApJ...728...30T} and {\it RHESSI} \citep{2009ApJ...704..863S,2009ApJ...704L..58R} to constrain the high temperature component. X-ray spectrometers are required to robustly diagnose this hottest emission but there have been few instruments capable of quiescent AR observations: a Bragg crystal on a Skylark sounding rocket \citep{1975SoPh...42..183P} and {\it SMM}'s Flat Crystal Spectrometer (FCS) \citep{1996ApJ...473..519S,2014A&A...565A..14D}. These show weak emission from lines with peak formation temperatures up to 10~MK but emission mainly over 3-5~MK, with very steeply falling DEMs above 5~MK \citep{2014A&A...565A..14D}. Observations above 2~keV typically have poorer spectral resolution but the bremsstrahlung continuum dominates. Temperature diagnostics from the continuum have the advantage of being insensitive to non-equilibrium ionization that can affect line measurements \citep[e.g.][]{2003A&A...401..699B}. Full-disc observations of various ``non-flaring'' ARs have found higher temperatures of up to 11~MK with a {\it SDO/EVE} sounding rocket's X123 \citep{2015ApJ...802L...2C}, 6.6~MK with {\it CORONAS-Photon}/SphinX \citep{2012A&A...544A.139M} and 6-8~MK with {\it RHESSI} \citep{2009ApJ...697...94M}.

\begin{figure*}
\centering
\includegraphics[width=1.8\columnwidth]{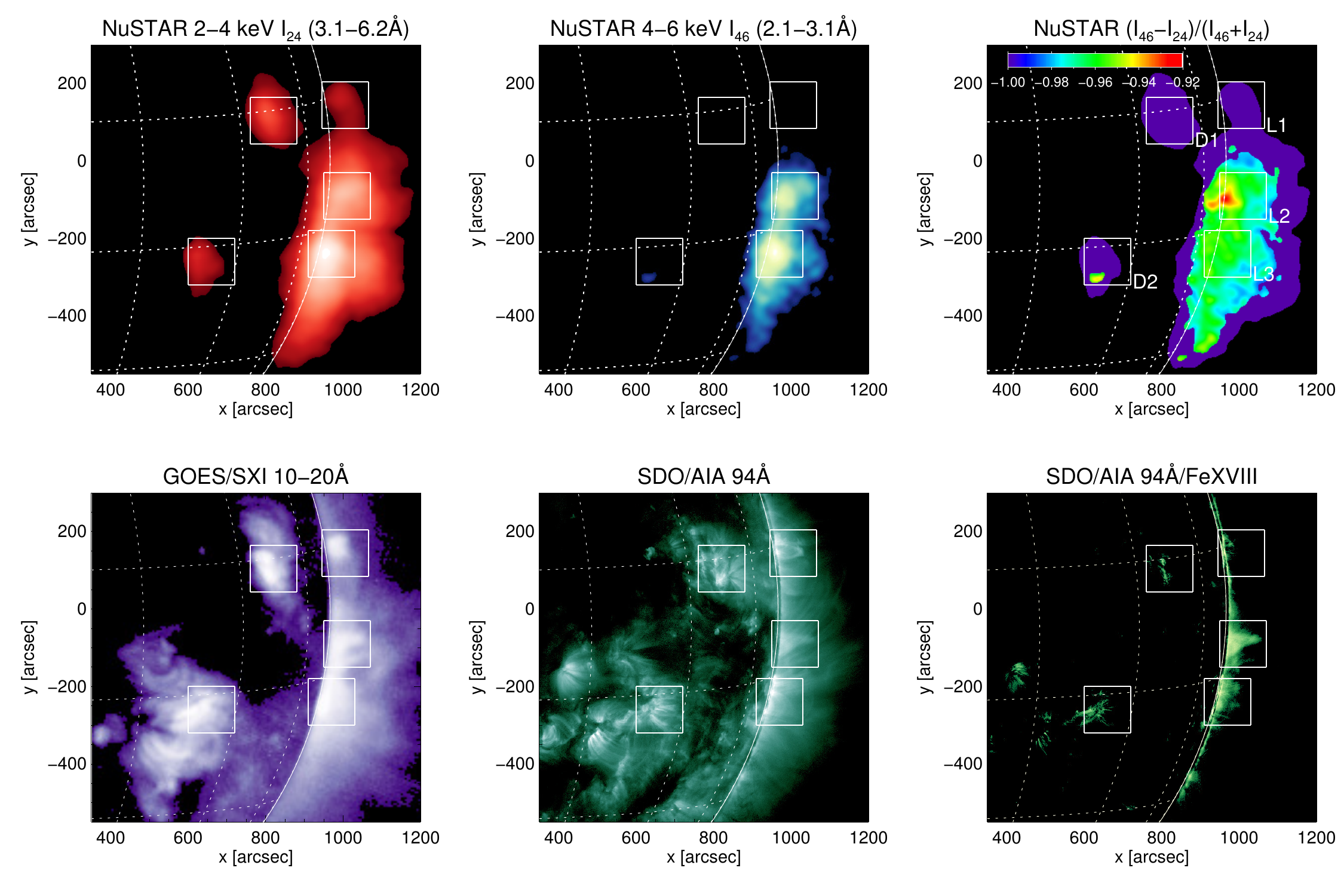}
\caption{(Top) {\it NuSTAR} X-ray images of the limb ARs in 2-4 and 4-6 keV, and the resulting differential hardness ratio between (top right). (Bottom) The lower energy SXR {\it GOES}/SXI and EUV {\it SDO}/AIA 94\AA~images of the same regions. Also shown is the {\it SDO}/AIA 94\AA~\ion{Fe}{18} component (bottom right). The boxes indicate the spectral analysis regions.}\label{fig:8img}
\end{figure*}

If nanoflares are just energetically smaller versions of flares/microflares then they should also release stored magnetic energy in the form of accelerated electrons, as even the smallest microflares ({\it GOES} A,B-Class flares) show hard X-ray (HXR) emission \citep{1984ApJ...283..421L,2008A&A...481L..45H}. Similar physics seems possible given that flare energies appear to behave as a power-law frequency distribution \citep[e.g.][]{2011SSRv..159..263H}, with the occurrence increasing with decreasing magnitude. Accelerated electrons are detectable in flares via their bremsstrahlung emission at higher X-ray energies ($>$10~keV) but for quiescent periods these signatures have remained elusive. {\it RHESSI} has been prolific at observing flare/microflare HXRs but struggles with quiescent emission due to its indirect imaging and high non-solar background. Only high-temperature emission was found for quiescent ARs using {\it RHESSI} \citep{2009ApJ...697...94M} and 3-200~keV upper limits for the quiet Sun \citep{2007ApJ...659L..77H,2010ApJ...724..487H}. The {\it FOXSI} sounding rocket has higher X-ray sensitivity and direct imaging and has briefly observed the Sun twice. It was able to constrain the high temperature emission from a non-flaring AR but has not detected anything non-thermal \citep{2014ApJ...793L..32K,2014PASJ...66S..15I}. The presence of accelerated electrons has been suggested from {\it IRIS} ultraviolet observation \citep{2014Sci...346B.315T}, as the rapid variability of intensities and velocities at hot coronal loop footpoints are consistent with simulations of heating by accelerated electron beams. However X-ray observations are needed to definitively detect non-thermal emission.

In this paper we present the first directly imaged and spectrally resolved X-rays above 2~keV from quiescent ARs using the {\it Nuclear Spectroscopic Telescope Array}  ({\it NuSTAR}; \citealt{2013ApJ...770..103H}). {\it NuSTAR} is the first focusing optics telescope that covers the HXR bandpass (2.5-78~keV) and has a higher sensitivity than {\it RHESSI} \citep{2002SoPh..210....3L}, with an effective area $20\times$ larger at $10$~keV and lower background. {\it NuSTAR} provides a unique X-ray view of the Sun but is not an optimized solar telescope. The solar pointings therefore present challenges, especially to use {\it NuSTAR}'s full sensitivity. We briefly discuss this in \S\ref{sec:ns_sum} and a full discussion is available in \citet{2016BWG}. In \S\ref{sec:ns_data} we present {\it NuSTAR} X-ray images and spectroscopy of several ARs observed on 2014 November 1, showing detection of emission at $3.1-4.4$~MK. We determine constraints to the emission above 5~MK, which are an order of magnitude more restrictive than previous observations, in \S\ref{sec:ht}.

\section{NuSTAR Solar Observations}\label{sec:ns_sum}

{\it NuSTAR} carries two identical co-aligned mirror modules that focus onto two focal-plane modules (FPMA, FPMB). Each  has a $12'\times 12'$ field of view (FoV) and FWHM of $\sim$18$''$ \citep{2015ApJS..220....8M}. The focal-plane modules are $2\times2$ arrays of CdZnTe detectors each with $32\times32$ pixels operated in a photon-counting mode. It has a useable energy-range of 2.5-78~keV with a resolution of $\sim 0.4$~keV FWHM below 20~keV. The effective area is well calibrated down to 3~keV, with only small deviations to 2.5~keV, becoming substantial at lower energies due to trigger threshold uncertainty \citep[see][]{2016BWG}. The readout time per event of 2.5 ms gives a maximum throughput of 400 counts sec$^{-1}$ per module. The data are processed using the {\it NuSTAR} Data Analysis software v1.5.1 and {\it NuSTAR} CALDB 20150414\footnote{\url{http://heasarc.gsfc.nasa.gov/docs/nustar/analysis/}}, generating an event list from which images/spectra are created. To reduce the effects of pile-up -- multiple low energy photons being recorded as a single high energy count \citep[e.g.][]{1975SSI.....1..389D} -- we restrict our analysis to single-pixel (``Grade 0'') events \citep{2016BWG}. {\it NuSTAR}'s optics are such that objects outside the FoV can add to the detected background via photons that reflect only once, instead of twice for properly focused photons, allowing them to reach the focal plane as ``ghost rays'' \citep{2015ApJS..220....8M}. These have a well-characterised shape and behaviour. Without detailed knowledge of the brightness and variability of the ghost ray sources, no corrections are possible. Optimal {\it NuSTAR} solar observations therefore require weak sources outside the target region.

The data in this paper are from the first solar science campaign on 2014 November 1 \citep{2016BWG}, consisting of four orbits of observations (each about an hour). We present the analysis of the fourth orbit as it provides the most stable view of ARs near the west limb and has the highest livetime, with effective exposures of about $3$ and $11$ seconds.

\begin{figure*}
\centering
\includegraphics[width=1.8\columnwidth]{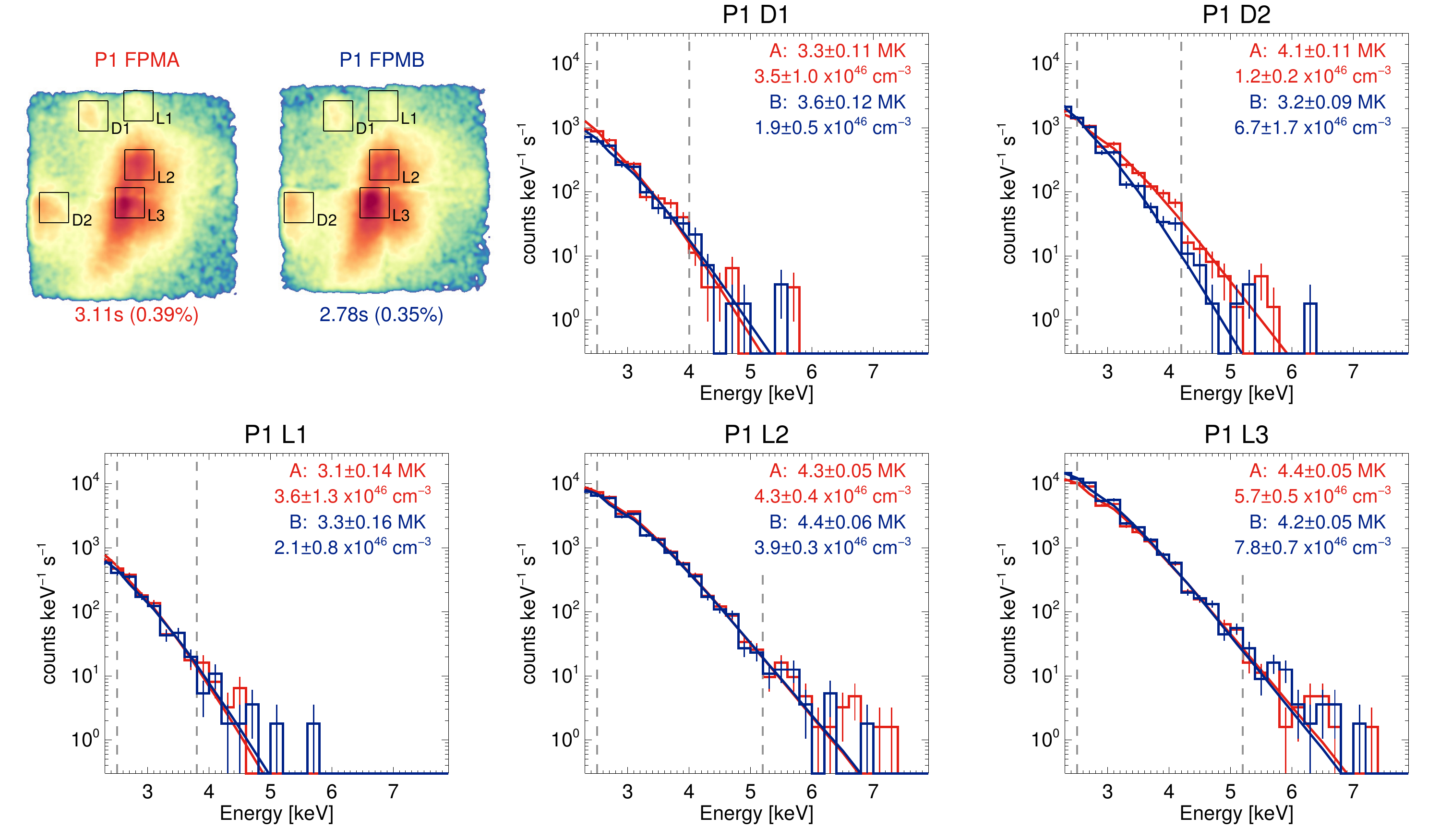}
\caption{{\it NuSTAR} spectra and isothermal fits (with 1$\sigma$ uncertainties) for regions D1, D2, L1, L2 and L3 during P1, from FPMA (red) and FPMB (blue). The vertical dashed lines indicate the energy range of the fit. The effective exposure times (and livetime percentage) are shown below each image.}
\label{fig:specsP1}
\end{figure*}

\begin{figure*}
\centering
\includegraphics[width=1.8\columnwidth]{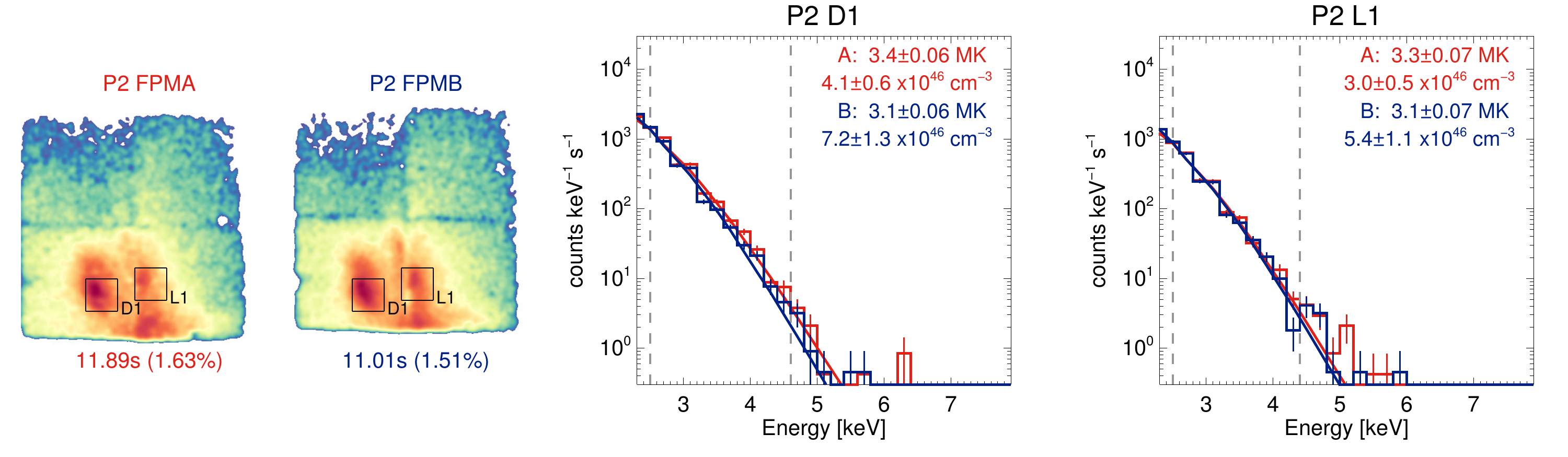}
\caption{{\it NuSTAR} spectra and isothermal fits (with 1$\sigma$ uncertainties) for regions D1 and L1 during P2, from FPMA (red) and FPMB (blue). The vertical dashed lines indicate the energy range of the fit. The effective exposure times (and livetime percentage) are shown below each image.}
\label{fig:specsP2}
\end{figure*}

\section{NuSTAR Quiescent ARs}\label{sec:ns_data}
\subsection{Time Profile and Image Mosaic}

Figure~\ref{fig:lght_mos} shows the fourth orbit observations starting 21:34~UT 2014 November 1, which consists of four pointings (P1, P2, P3, and P4) going from solar W~to N. For the majority of this orbit the solar position was found by a single combination of star trackers, or Camera Head Units (CHUs; \citealt{2015Walton}). A given CHU combination is accurate to $<$1.5$'$ \citep{2016BWG} and varies with the pointing. The {\it NuSTAR} images from P1 and P2 are individually shifted to match the AR locations from the EUV/SXR images (see Figure \ref{fig:8img}). As there are no clear features in P3, the P2 shift was used. For P4 there are no sources and it is a different CHU combination, so data from the previous orbit during the same combination, while observing the ARs, were used for co-alignment.

The time series in Figure~\ref{fig:lght_mos} compares the {\it NuSTAR} livetime percentage (averaged over FPMA and FPMB, individually similar) with the {\it RHESSI} and {\it GOES} observations. The {\it NuSTAR} livetime is very low (about 0.4$\%$) and near constant during P1, which is due to the quiescent ARs. These regions can be seen in the $2-78$~keV image in Figure~\ref{fig:lght_mos}, from FPMA and FPMB combined. In P2, the livetime increases due to fewer ARs in the FoV and increases further in P3 and P4 once there are no ARs in the FoV. In these pointings the background (blue-green in Figure~\ref{fig:lght_mos}) is ghost rays from ARs on the disk (the imaged limb ones and those elsewhere on the disk). The cross shape, most evident in P4, is from the gap between the detector quadrants. During P4 a small microflare occured outside {\it NuSTAR}'s FoV, detected by {\it GOES} and {\it RHESSI}, appearing as a dip in the {\it NuSTAR} livetime due to the increase in ghost rays. During the times when {\it NuSTAR} was directly imaging the limb ARs, the full disc {\it GOES} flux was $\sim 4\times10^{-7}$~Wm$^{-2}$, mostly coming from ARs outside {\it NuSTAR}'s FoV. These ghost rays are about $100\times$ weaker than the directly imaged ARs \citep[see][]{2016BWG} and so will have a negligible contribution.

\subsection{X-ray and EUV Image Comparison}

Combining the P1 and P2 data we produce 2-4 keV and 4-6 keV livetime-corrected images (with $\sim$7$''$ Gaussian smoothing), and compare these to {\it SDO}/AIA EUV and {\it GOES}/SXI soft X-ray (SXR) images (Figure~\ref{fig:8img}). The regions in EUV/SXR show little variability during the {\it NuSTAR} observation time, and those shown are averaged over that period. 

In the {\it NuSTAR} 2-4 keV image there are five distinct regions which match features in the EUV/SXR images. They correspond to (and we label as) NOAA AR12195 (D1), AR12196 (D2), and above-the-limb sources (L1), highly occulted AR12192 (L2) and AR12194 (L3). The major region AR12192 produced several X-class flares when on the disk and contained expansive coronal loops, still visible above the limb in EUV/SXR. {\it SDO}/AIA 94\AA's temperature response contains cooler (0.5-1~MK) and hotter (3-10~MK) components \citep{2012ApJ...745..111T,2014SoPh..289.2377B}. So for comparison to the {\it NuSTAR} images we isolate the emission above 3~MK, from \ion{Fe}{18} \citep{2011ApJ...734...90W,2013A&A...558A..73D} (bottom right Figure~\ref{fig:8img}). These show that the hottest emission is from more compact regions, the brightest from L2 and L3. This agrees with the {\it NuSTAR} 4-6 keV image, which only shows discernible emission from these locations. We also show the differential hardness ratio $(I_{46}-I_{24})/(I_{46}+I_{24})$ of the 4-6 keV ($I_{46}$) to 2-4 keV ($I_{24}$) emission, finding the highest spectral hardness in L2 and L3. 

\subsection{X-ray Imaging Spectroscopy}

For each AR we accumulate the {\it NuSTAR} spectrum over a $120''\times 120''$ region (Figures~\ref{fig:specsP1} and \ref{fig:specsP2}). This is done separately for P1, P2 and FPMA, FPMB as the instrument response is different for each. These spectra are binned with 0.2~keV resolution and only include Grade 0 (single-pixel hit, minimizing pile-up) events, in the dominant CHU combination. To forward-fit a model to this data using SolarSoft/OSPEX\footnote{\url{http://hesperia.gsfc.nasa.gov/ssw/packages/spex/doc/}} we need a spectral response matrix (SRM) for each region, generated from the Redistribution Matrix and Ancillary Response Files (RMF, ARF) produced by the {\it NuSTAR} Data Analysis Software\footnote{\url{http://heasarc.gsfc.nasa.gov/docs/nustar/analysis/}}.

Figure~\ref{fig:specsP1} and \ref{fig:specsP2} shows forward-fits of CHIANTI 7.1 \citep{1997A&AS..125..149D,2013ApJ...763...86L} isothermal models to each {\it NuSTAR} spectrum. We fit from 2.5~keV (the minimum useable energy) to the highest energy with $\sim 10$ counts per bin, so that the uncertainties are Gaussian (as OSPEX uses the chi-square test). These isothermal models fit the data well, with the few excess counts at higher energies consistent within Poisson statistics. Similar temperatures and emission measures are achieved for each region in FPMA and FPMB except for D2 (top right Figure~\ref{fig:specsP1}). Here the FPMB spectrum is from the detector edge, where the ARF is poorly calibrated \citep{2015ApJS..220....8M} and there might be missing counts. The regions observed in both P1 and P2 show similar values for each fit, within the relative calibration \citep{2016BWG}. The P2 values are more robust as the regions are closer to the imaging axis and observed with a higher effective exposure (about 11~s instead of 3~s). 

We can compare the {\it NuSTAR} isothermal fits to the {\it SDO}/AIA observations of the regions by folding them through the 94\AA~temperature response \citep{2014SoPh..289.2377B}. We find that the {\it NuSTAR} fits reproduces 3-17\% of the observed 94\AA~flux and 10-82\% of the 94\AA~\ion{Fe}{18}~flux, consistent with {\it NuSTAR} only observing part of the multi-thermal emission seen by {\it SDO}/AIA. This is due to the weak response of {\it NuSTAR} to cooler temperatures and the short effective exposure times (limiting the dynamic range of the spectra at higher energies). To improve our sensitivity to a faint high-temperature (or non-thermal) component we could increase our exposure by observing for longer (more than 790s) and/or during times of weaker solar activity (and hence achieve livetimes larger than 0.4\% and 1.6\%).

\begin{figure}
\centering
\includegraphics[width=0.9\columnwidth]{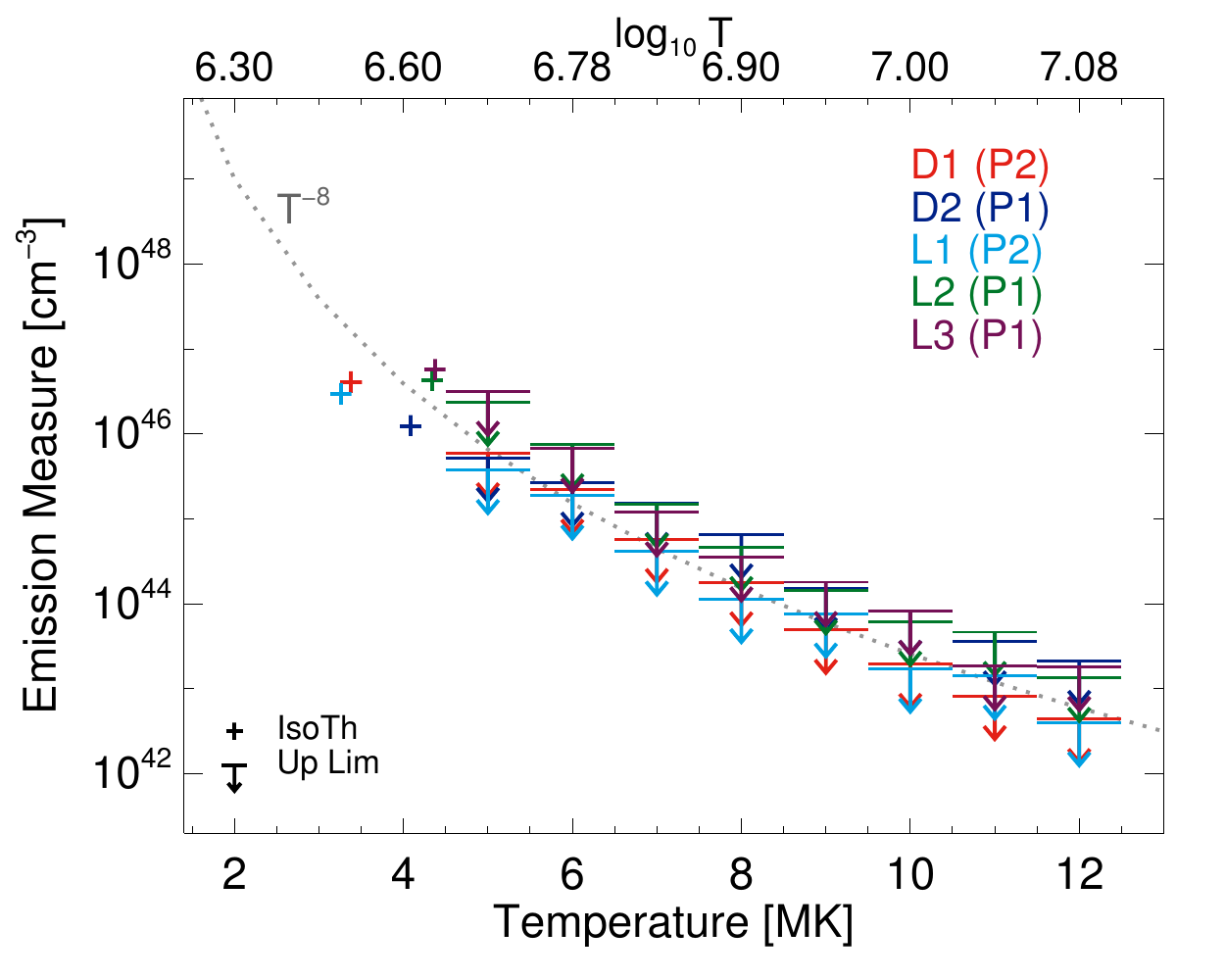}\\
\includegraphics[width=0.9\columnwidth]{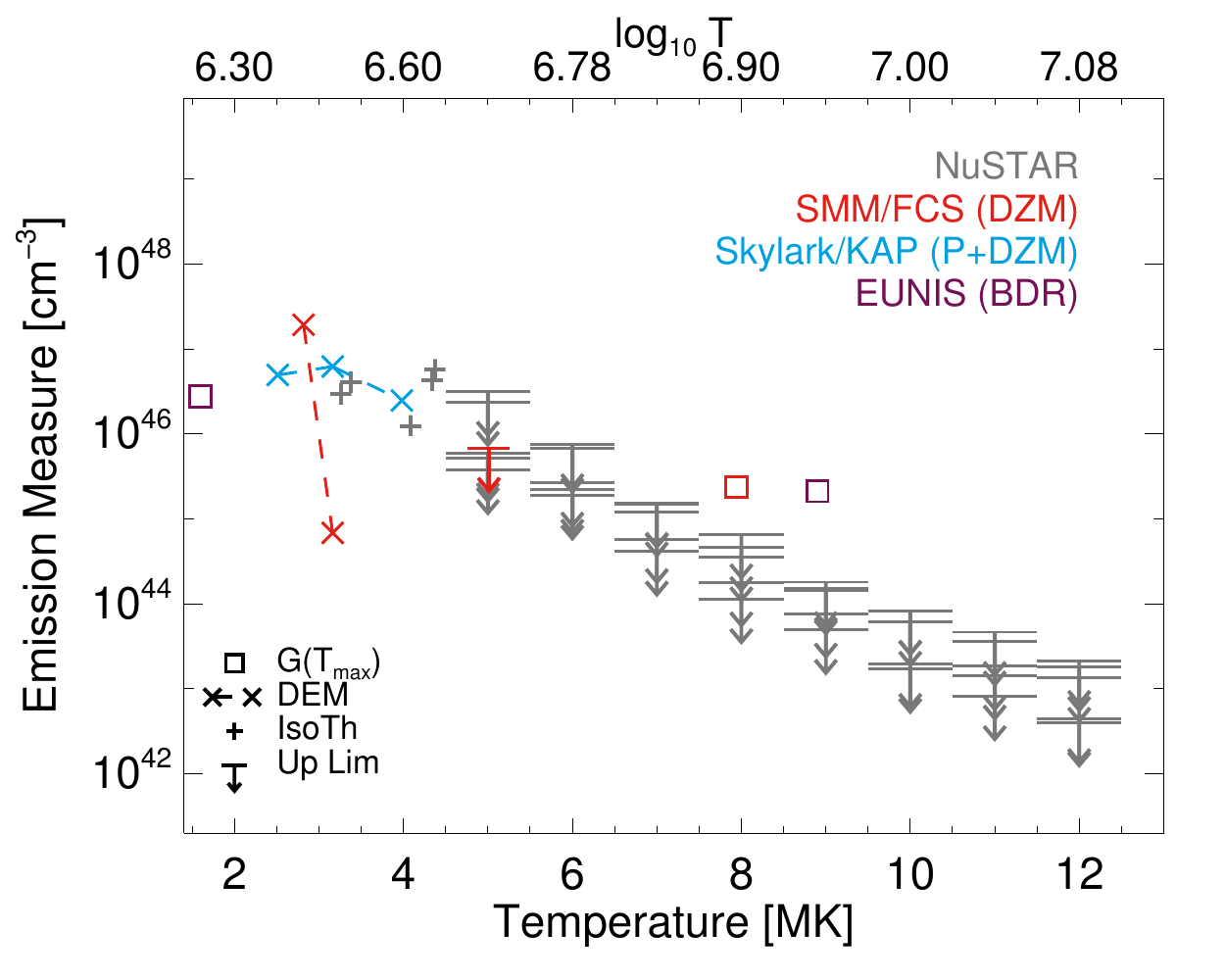}\\
\includegraphics[width=0.9\columnwidth]{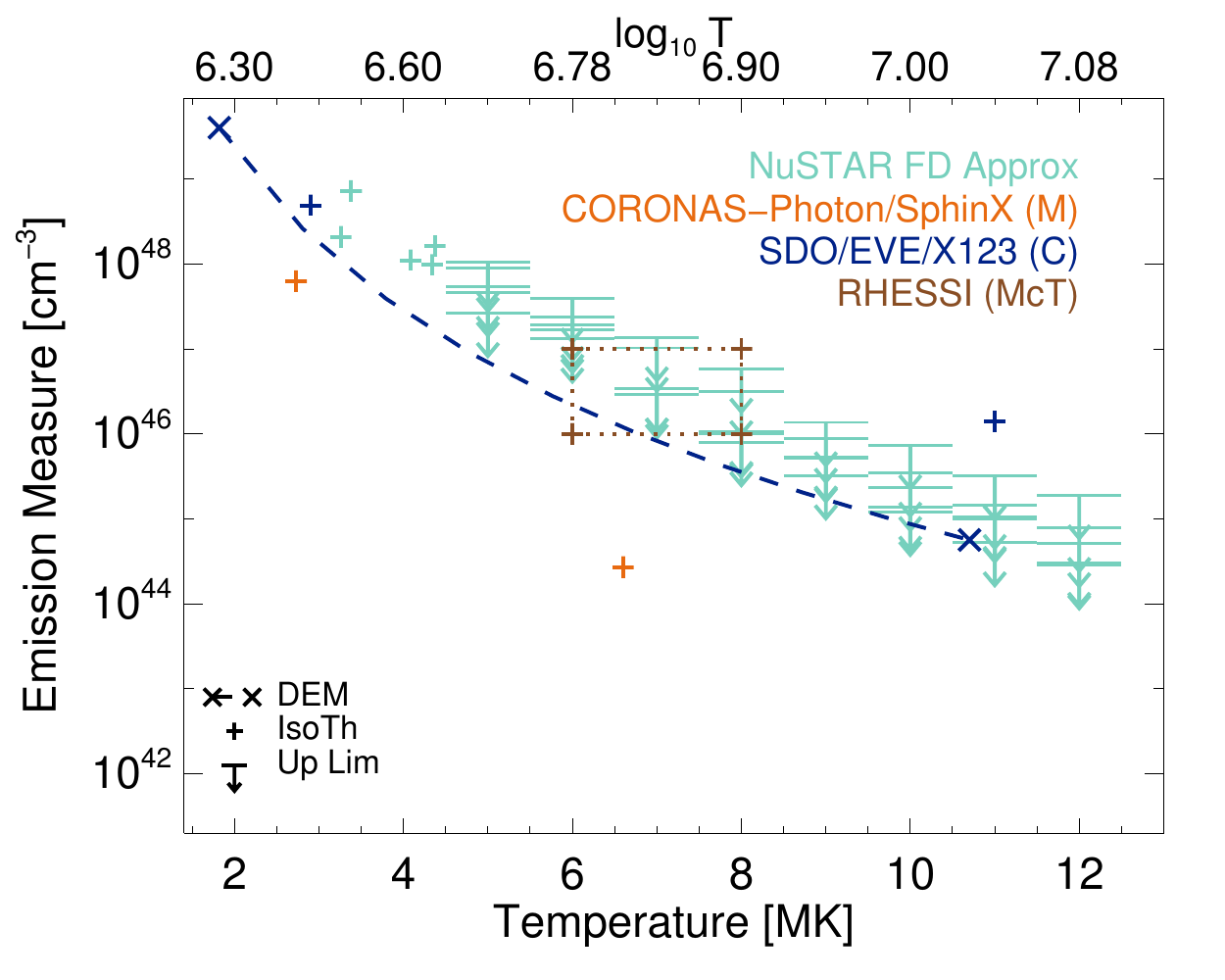}\\
\caption{(Top) Upper limits to the emission from a second thermal component for each of the {\it NuSTAR} region spectra. For comparison the dotted grey lines shows $EM\propto T^{-8}$. (Middle) The {\it NuSTAR} high-temperature upper limits (gray) compared to EUV (\citealt{2014ApJ...790..112B} - BDR) and SXR (\citealt{1975SoPh...42..183P} - P+DZM; \citealt{2014A&A...565A..14D} - DZM) spectroscopy. (Bottom) The {\it NuSTAR} high-temperature upper limits scaled to full-disk emission (turquoise), compared to other full-disk X-ray observations 
(\citealt{2012A&A...544A.139M} - M; \citealt{2015ApJ...802L...2C} - C; \citealt{2009ApJ...697...94M} - McT). The symbols indicate the different ways in which the values were obtained: square for maximum of $G(T)$, cross and dashed line for DEM fit, plus for isothermal fit, arrow for upper limit.}
\label{fig:mc_ht}
\end{figure}

\section{High Temperature Constraints}\label{sec:ht}

The {\it NuSTAR} spectra do not show additional high temperature ($>$5~MK) or non-thermal emission, but we can calculate upper limits on this emission. We concentrate on the possible high temperature constraints as we know that hot plasma could be present (from EUV/SXR observations) and would have to be accounted for before non-thermal constraints are attempted. 

We determine the upper limits on the emission of an additional hotter component for all regions using FPMA (which has a slightly higher livetime) and P1 for D2, L2 and L3 and P2 for D1 and L1. A Monte Carlo approach is used with the livetime and SRM of each region to generate a synthetic {\it NuSTAR} spectrum of a two-component thermal model (one using the fitted thermal parameters, the other a chosen temperature between $5-12$ MK). The emission measure of the second component starts with a large value (that of the lower temperature fit) and is iteratively reduced until there are fewer than 4 counts above 6~keV, consistent within $2\sigma$ of a null detection \citep{1986ApJ...303..336G}. This is repeated for each temperature and region. The resulting upper limits (Figure~\ref{fig:mc_ht}) range from about $10^{46}$cm$^{-3}$ at 5~MK to $10^{43}$cm$^{-3}$ at 12~MK, with the lowest limits coming from the observations with the largest effective exposure, P2.

Compared to the observations from {\it EUNIS} \citep{2014ApJ...790..112B} and {\it SMM}/FCS \citep{2014A&A...565A..14D} the {\it NuSTAR} limits are at least an order of magnitude lower (Figure~\ref{fig:mc_ht}, middle panel). This might be due to these previous studies observing ARs with brighter hot emission. Even within the {\it NuSTAR} limits there is about an order of magnitude spread from the different regions. The high temperature values from {\it EUNIS} and {\it SMM}/FCS are calculated using the maximum of the contribution functions $G(T)$, an isothermal approach using the peak formation temperature. If the emission is due to a wider range of temperatures, and the DEM is expected to be sharply falling with temperature, then the actual emission is from lower temperatures. This ``effective temperature'' was calculated for {\it SMM}/FCS \ion{Fe}{18} \citep{2014A&A...565A..14D} giving emission at 5~MK, instead of the 8~MK using the peak formation temperature (red upper limit versus square in Figure~\ref{fig:mc_ht}, middle panel). The DEMs of quiescent ARs from X-ray spectroscopy \citep{1975SoPh...42..183P,2014A&A...565A..14D} show emission over 2-4~MK. The {\it NuSTAR} isothermal fits are consistent with the {\it Skylark} results but higher than the {\it SMM}/FCS. This again could be indicative of quiescent ARs producing a wide range of emission.

We scale the {\it NuSTAR} limits by the fraction of the {\it SDO}/AIA 94\AA~\ion{Fe}{18} emission in each region to compare to other full-disk X-ray spectroscopy of quiescent ARs. About 0.5\% of the full-disk emission comes from D1, 1.1\% for D2, 1.4\% for L1, 4.4\% for L2 and 3.5\% for L3. These {\it NuSTAR} full disk limits (Figure~\ref{fig:mc_ht}, bottom panel) match the range of emission observed with {\it RHESSI} \citep{2009ApJ...697...94M} and the DEM from X123 \citep{2015ApJ...802L...2C}. The SphinX two-thermal fit gives emission considerably lower than the scaled {\it NuSTAR} values, which is still consistent as they are upper limits. The SphinX values are small as they are from a period of very low solar activity and averaged over a month of observations.

\section{Conclusions}\label{sec:discuss}

{\it NuSTAR} is a uniquely sensitive telescope, capable of observing faint X-ray emission from the non-flaring Sun. This paper shows for the first time X-ray emission above $2$~keV directly imaged from quiescent ARs. The spectra of these regions are well fitted by 3.1-4.4~MK isothermal emission. We have not detected a higher temperature or non-thermal contribution to the HXR spectra for these ARs. We place strict constraints on hotter sources, requiring them to decrease with at least $T^{-8}$, which is consistent with impulsive heating models \citep[e.g.][]{1994ApJ...422..381C,2008ApJ...682.1351K}. To increase {\it NuSTAR}`s spectral dynamic range, improving our ability to detect or further constrain high temperature and non-thermal contributions, we need observations with larger effective exposure times. This can be achieved by longer duration observations and/or diminishing solar activity (producing lower deadtime). Accessing {\it NuSTAR}`s full sensitivity, combined with other new data from the {\it FOXSI} \citep{2014ApJ...793L..32K} and {\it MAGIXS} \citep{2011SPIE.8147E..1MK} sounding rockets, will provide crucial steps towards improved X-ray observations of the Sun and understanding the nature of its quiescent energy release.

\acknowledgments
This paper made use of data from the NuSTAR mission, a project led by the California Institute of Technology, managed by the Jet Propulsion Laboratory, funded by the National Aeronautics and Space Administration. We thank the NuSTAR Operations, Software and Calibration teams for support with the execution and analysis of these observations. This research made use of the NuSTAR Data Analysis Software (NUSTARDAS) jointly developed by the ASI Science Data Center (ASDC, Italy), and the California Institute of Technology (USA). This work is supported by: NASA grants NNX12AJ36G, NNX14AG07G, IGH (Royal Society University Research Fellowship), AM (NASA Earth Space Science Fellowship, NNX13AM41H), SK (Swiss National Science Foundation, 200021-140308), AC (NASA NNX15AK26G, NNX14AN84G). Thanks to Kim Tolbert for OSPEX help.

\end{document}